\documentclass[aps,prl,twocolumn,superscriptaddress]{revtex4-1}

\usepackage{graphicx}
\usepackage[rightcaption]{sidecap}
\usepackage{floatrow}
\usepackage{amsmath}
\usepackage{array}
\usepackage{color}
\usepackage{xfrac}
\newcolumntype{@}{>{\global\let\currentrowstyle\relax}}
\newcolumntype{^}{>{\currentrowstyle}}

\usepackage{comment}

\newcommand{\sState}{$^2\text{S}_{1/2}\,$}
\newcommand{\pState}{$^2\text{P}_{1/2}\,$}
\newcommand{\PState}{$^2\text{P}_{3/2}\,$}
\newcommand{\dState}{$^2\text{D}_{3/2}\,$}
\newcommand{\DState}{$^2\text{D}_{5/2}\,$}

 \newcommand{\ba}{$^{133}\text{Ba}^+\,$}
\newcommand{\natba}{$^{138}\text{Ba}^+\,$}

\newcommand{\zero}{$\vert 0 \rangle\,$}
\newcommand{\one}{$\vert 1 \rangle\,$}
\newcommand{\two}{$\vert 2 \rangle\,$}
\newcommand{\ket}[1]{{\left| {#1} \right\rangle}}

\begin{document}


\title{High fidelity manipulation of a qubit built from a synthetic nucleus}


\author{Justin E. Christensen} \affiliation{Department of Physics and Astronomy, University of California -- Los Angeles, Los Angeles, California, 90095, USA}
\author{David Hucul} \affiliation{Department of Physics and Astronomy, University of California -- Los Angeles, Los Angeles, California, 90095, USA} \affiliation{United States Air Force Research Laboratory, Rome, New York, 13441, USA}
\author{Wesley C. Campbell}
\affiliation{Department of Physics and Astronomy, University of California -- Los Angeles, Los Angeles, California, 90095, USA}
\affiliation{UCLA Center for Quantum Science and Engineering, University of California -- Los Angeles, Los Angeles, California 90095, USA}
\author{Eric R. Hudson}
\affiliation{Department of Physics and Astronomy, University of California -- Los Angeles, Los Angeles, California, 90095, USA}
\affiliation{UCLA Center for Quantum Science and Engineering, University of California -- Los Angeles, Los Angeles, California 90095, USA}


\date{\today}

\begin{abstract}
The recently demonstrated trapping and laser cooling of \ba \ has opened the door to the use of 
this nearly ideal atom for quantum information processing. 
However, before high fidelity qubit operations can be performed, a number of unknown state energies are needed. 
Here, we report measurements of the \PState\ and \DState\ hyperfine splittings, as well as the 
\PState\ $\leftrightarrow$ \sState\ and \PState\ $\leftrightarrow$ \DState\ transition frequencies. 
Using these transitions, we demonstrate high fidelity \ba \ hyperfine qubit manipulation with electron shelving detection to benchmark qubit state preparation and measurement (SPAM). 
Using single-shot, threshold discrimination, we measure an average SPAM fidelity of $\mathcal{F} = 0.99971(6)$, a factor of $\approx$ 2 improvement over the best reported performance of any qubit.

\end{abstract}


\maketitle

Quantum error correction allows an imperfect quantum computer to perform reliable calculations beyond the capability of classical computers~\cite{steane:1998, preskill:1997,gottesman:2009}.
However, even with the lowest reported error rates \cite{gaebler:2016, hume:2007, jeffrey:2014, harty:2014, ballance:2016, bermudez:2017, ernhard:2019,Wu:2019}, the number of qubits ($N_q$) required to achieve fault tolerance is projected \cite{fowler:2012} to be  \emph{significantly} larger than the state of the art \cite{ionq:2019, friis:2018, debnath:2016,neill:2018}.
Nonetheless, noisy intermediate-scale quantum (NISQ) devices \cite{preskill:2018} are currently being employed to tackle important problems without fault tolerance \cite{nam:2019, zhang:2017, hempel:2018, gorman:2018, landsman:2019,hucul:2015,kokail:2019}.

For these NISQ devices, state preparation and measurement (SPAM) infidelity ($\epsilon_s$) causes a reduction in computational fidelity that is exponential in qubit number, $\mathcal{F}_s = (1-\epsilon_{s})^{N_q}$.
The requirement to perform faithful SPAM therefore limits the number of qubits to $N_q < \ln(2)/\epsilon_s$. 
While state readout error correction techniques can effectively lower measurement infidelity, they generally require a number of measurements that grows exponentially with $N_q$ and single-shot readout infidelity \cite{shen:2012}.
For these reasons, and given the desire to increase $N_q$ to tackle problems beyond the reach of classical computers, it is therefore important to develop new means to improve $\epsilon_s$. 

The A = 133 isotope of barium provides a potential path to improving fidelities in atomic ion quantum computing, as this isotope combines the advantages of many different ion qubits into a single system \cite{hucul:2017}. 
\ba \ has nuclear spin $I = 1/2$, which as we show here, allows fast, robust state preparation and readout of the hyperfine qubit.
It possesses both $m_F = 0$ hyperfine and optical ``clock" state qubits, which are relatively insensitive to magnetic fields ($m_F$ is the projection quantum number of the total angular momentum $F$)~\cite{wang:2017}. 
It also possesses metastable $D$ states ($\tau \sim$ 1 min), allowing high fidelity readout, and long-wavelength transitions enabling the use of photonic technologies developed for the visible and near-infrared spectrum. 
However, before these advantages can be realized, a number of unknown hyperfine and electronic transition frequencies must be determined. 

Here, we measure the previously unknown \PState\ and \DState\ hyperfine structure, as well as the \PState $\leftrightarrow$ \sState\ and \PState $\leftrightarrow$ \DState \ electronic transition frequencies. 
Using this knowledge, we demonstrate \ba\ hyperfine qubit manipulation and electron shelving detection.
Using a simple threshold discrimination and modest fluorescence  collection optics (0.28 NA), we measure an average single-shot SPAM fidelity of $\mathcal{F} = 0.99971(6)$, the highest reported for any qubit.


In what follows, we first present qubit SPAM using standard hyperfine-selective optical cycling \cite{olmschenk:2007, Acton2006} combined with arbitrary qubit  rotations and a composite pulse sequence for high-fidelity state transfer. 
We then present measurement of the unknown hyperfine and electronic transition frequencies. Finally, we use this information to demonstrate high fidelity SPAM using electron shelving.

\begin{figure}[t!]
\includegraphics[width=0.99\columnwidth]{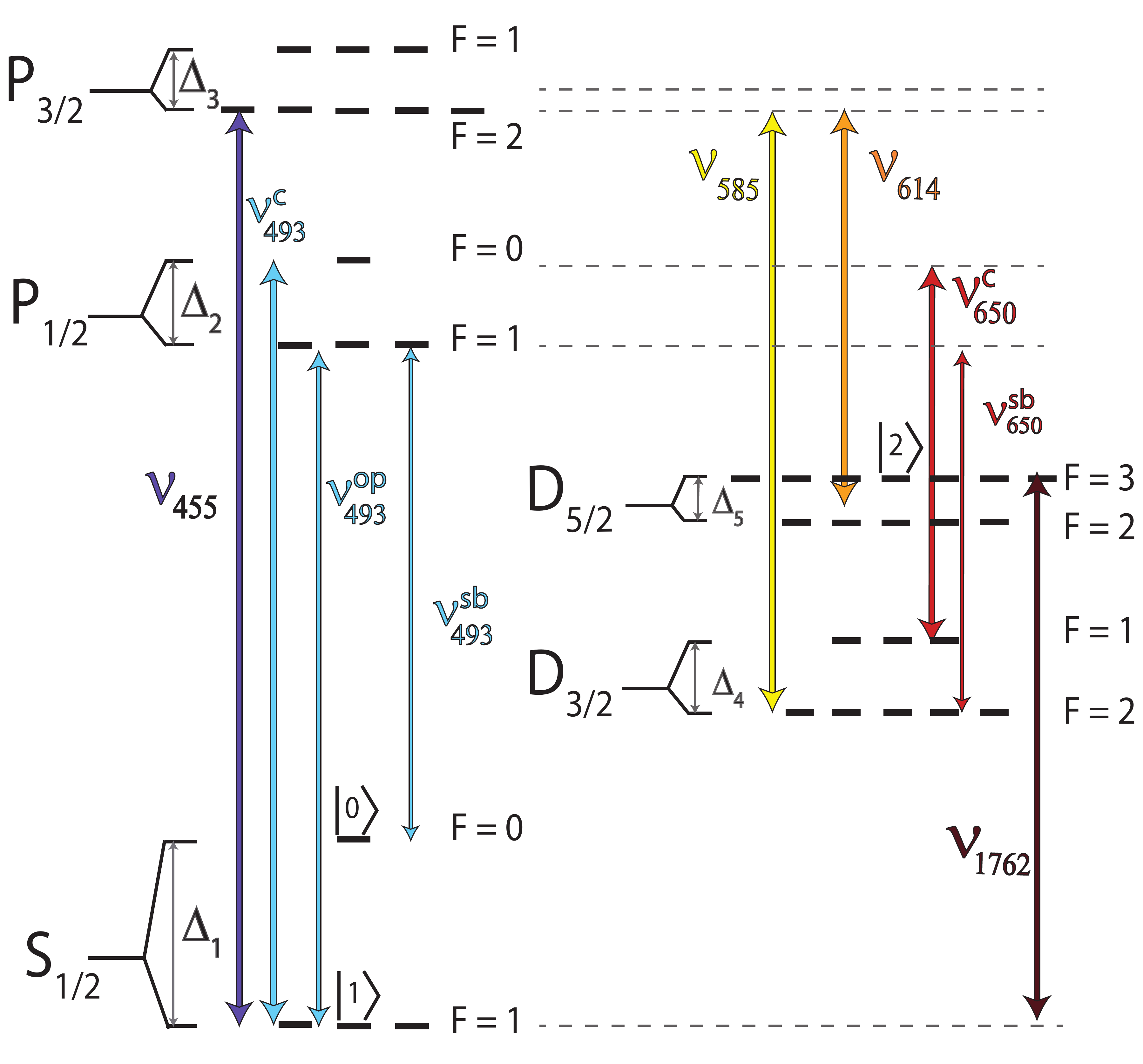}
\caption{\label{fig:ba_shelve}\ba energy level diagram and lasers required for laser cooling and high fidelity SPAM. Laser cooling is accomplished using lasers near 493 nm and 650 nm ($\nu_{493}^c$, $\nu_{650}^c$), and fiber EOMs for repumping sidebands ($\nu_{493}^{sb}$, $\nu_{493}^{op}$ and $\nu_{650}^{sb}$). The \zero\ state is initialized by removing $\nu_{493}^{sb}$ and adding $\nu_{493}^{op}$ with the 493 nm fiber EOM. Microwave radiation at $\Delta_1$ $\approx$ 9.925 GHz allows for arbitrary rotations on the Bloch sphere. Electron shelving of the \one \ state to the metastable \DState \ state is accomplished with a laser near 455 nm, 585 nm, and 650 nm ($\nu_{455}$, $\nu_{585}$, and $\nu_{650}$). A laser near 614 nm ($\nu_{614}$) is used to depopulate the \DState \ manifold after state detection. In future experiments a laser near 1762 nm ($\nu_{1762}$) can be used to directly manipulate the optical qubit clock-states \one \ and \two.} 
\end{figure}


We trap and laser cool \ba ions as described in \cite{hucul:2017}. Briefly, \ba ions are loaded into a linear Paul trap ($\omega \approx 2\pi \times 100$~kHz) by laser ablating an enriched BaCl$_2$ salt deposited on a platinum substrate. 
Laser cooling is accomplished using external cavity diode lasers (ECDLs) near 493 nm and 650 nm tuned to frequencies $\nu_{493}^c$ and $\nu_{650}^c$, with fiber electro-optic modulators (EOMs) used to add repumping sidebands $\nu_{493}^{sb}$, $\nu_{493}^{op}$, and $\nu_{650}^{sb}$ (Fig. \ref{fig:ba_shelve}). 

\begin{figure*}[th!]
\includegraphics[width=\columnwidth]{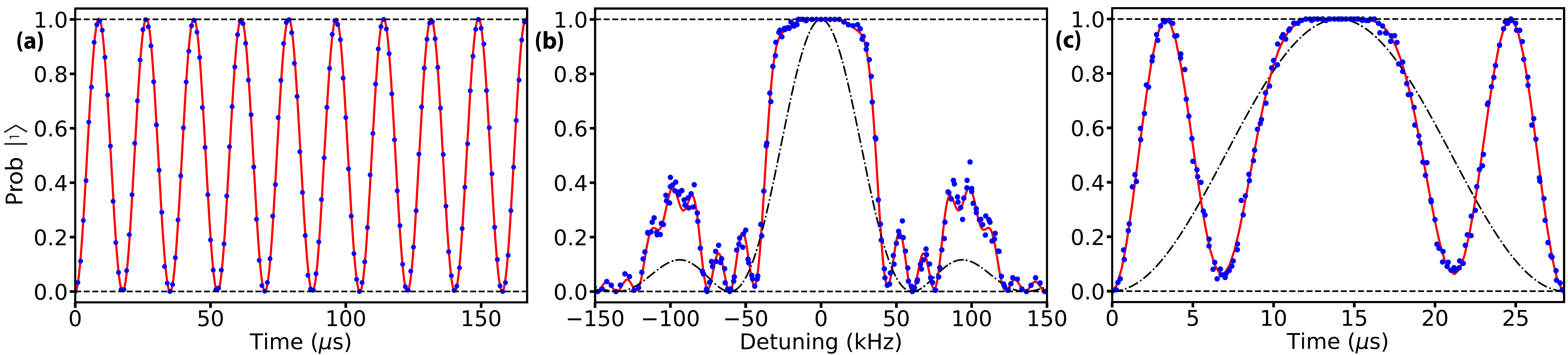}
\caption{\label{fig:knill} Microwaves near 9.925 GHz allow for arbitrary rotations $R(\theta,\phi)$ on the qubit Bloch sphere. (a) Probability of shelving \one\ after microwave rotations of the form $R(\Omega_Rt,0)$, where
$\Omega_R \approx 2\pi \times$ 57 kHz. To prepare the \one \ state, the five $\pi$-pulse CP Robust 180 sequence $R(\pi,\frac{\pi}{6})R(\pi,0)R(\pi,\frac{\pi}{2})R(\pi,0)R(\pi,\frac{\pi}{6})$  transfers population from the initially prepared \zero \ state. (b) Probability of shelving \one \ \emph{vs.} microwave detuning using the CP Robust 180 sequence with $\Omega_R \approx 2\pi \times$ 35 kHz. Points are experimental data and solid line represents theoretical prediction for this composite pulse sequence with no fit parameters. Dashed dotted line represents theory for single $\pi$-pulse, $R(\pi,0)$. (c) Pulse area ($t = \frac{\theta}{\Omega_R}$) scan at zero detuning using the CP Robust 180 sequence.}
\end{figure*}

The qubit is defined on the pair of $m_F$ = 0 ``clock" states in the ground state \sState manifold as \zero\ $\equiv \vert F \!\!=\!\! 0\rangle$ and \one\ $\equiv \vert F \!\!=\!\! 1; m_F \!\!=\!\! 0\rangle$.
This hyperfine qubit is initialized to the \zero \ state after Doppler cooling via optical pumping by applying frequencies $\nu_{493}^c$, $\nu_{493}^{op}$, $\nu_{650}^c$, and $\nu_{650}^{sb}$ (Fig. \ref{fig:ba_shelve}).
Rotations of the qubit Bloch vector about $\cos(\phi)\hat{x} + \sin(\phi)\hat{y}$ through angle $\theta$, $R(\theta,\phi)$, are accomplished by using microwave radiation near 9.925 GHz \cite{knab:1987} controlled by a modular digital synthesis platform \cite{pruttivarasin:2015}. 
An example rotation of the form $R(\Omega_{R} t,0)$ is shown in Fig.~\ref{fig:knill}(a), where the average population in state $\ket{1}$ found in 200 trials, measured with a technique described later, is plotted versus the application duration of microwave radiation at Rabi frequency $\Omega_R \approx 2\pi\times57$~kHz. 
The $\ket{1}$ state can be prepared after initialization into $\ket{0}$ by $R(\pi,0)$, however, we employ a composite pulse sequence, referred to as the CP Robust 180 sequence (attributed to E. Knill)~\cite{ryan:2010}, consisting of the five $\pi$-pulses $R(\pi,\frac{\pi}{6})R(\pi,0)R(\pi,\frac{\pi}{2})R(\pi,0)R(\pi,\frac{\pi}{6})$. 
As shown in Figs.~\ref{fig:knill}(b-c), the broad flat features in both curves near zero detuning and $\theta = \pi$ demonstrate resiliency to both pulse area and detuning errors as compared to single $\pi$-pulses, enabling robust day-to-day operation.

Typically for nuclear spin-1/2 hyperfine qubits, state readout is accomplished via hyperfine-selective optical cycling ($\nu_{493}^c$ and $\nu_{650}^c$ in Fig. \ref{fig:ba_shelve}) and collecting any resulting fluorescence. 
Projection into the \zero\ or \one\ state is then determined by e.g. a threshold discrimination on the number of collected photons, as an atom in the \one\ state scatters many photons, while an atom in the \zero\ state does not~\cite{Acton2006,olmschenk:2007}.
Using this hyperfine-selective optical cycling for SPAM, we measure the fraction of events in which an ion prepared in the \zero \ state was determined to be \one, $\epsilon_{\vert 0\rangle} = 3.03(4)  \times 10^{-2}$, and the fraction of experiments in which an ion prepared in the \one \ state was determined to be \zero, $\epsilon_{\vert 1 \rangle} = 8.65(9) \times 10^{-2}$. 
The average SPAM fidelity is defined as $\mathcal{F} = 1-\epsilon =1- \frac{1}{2}(\epsilon_{\vert 0\rangle} + \epsilon_{\vert 1 \rangle}) =  0.941(1)$. 
The fidelity of this technique is limited by off-resonant excitation to the $\vert$\pState$, F\!\! =\!\! 1\rangle$ manifold during the readout phase, which can decay to either \zero\ or $\ket{1}$, thereby causing misidentification of the original qubit state~\cite{olmschenk:2007}.

For high fidelity SPAM, \ba\ offers another, unique path to state detection. 
The \one\ qubit state can be shelved \cite{dehmelt:1975} to the long-lived ($\tau \approx$ 30 s) metastable \DState\ state via the \DState $\leftrightarrow$ \sState \ transition, or optically pumped via the \PState\ state (Fig. \ref{fig:ba_shelve}), followed by Doppler cooling for state readout. 
Projection into the \zero\ or \one\ state is then determined by e.g. a threshold discrimination on the number of collected photons, as an atom in the \zero\ state scatters many photons, while an atom in the \DState\ state, indicating \one, does not. 
Off-resonant scatter is negligible in this case as the Doppler cooling lasers are detuned by many THz from any \DState\ state transitions. 

\begin{figure}[t!]
\includegraphics[width=0.95\columnwidth]{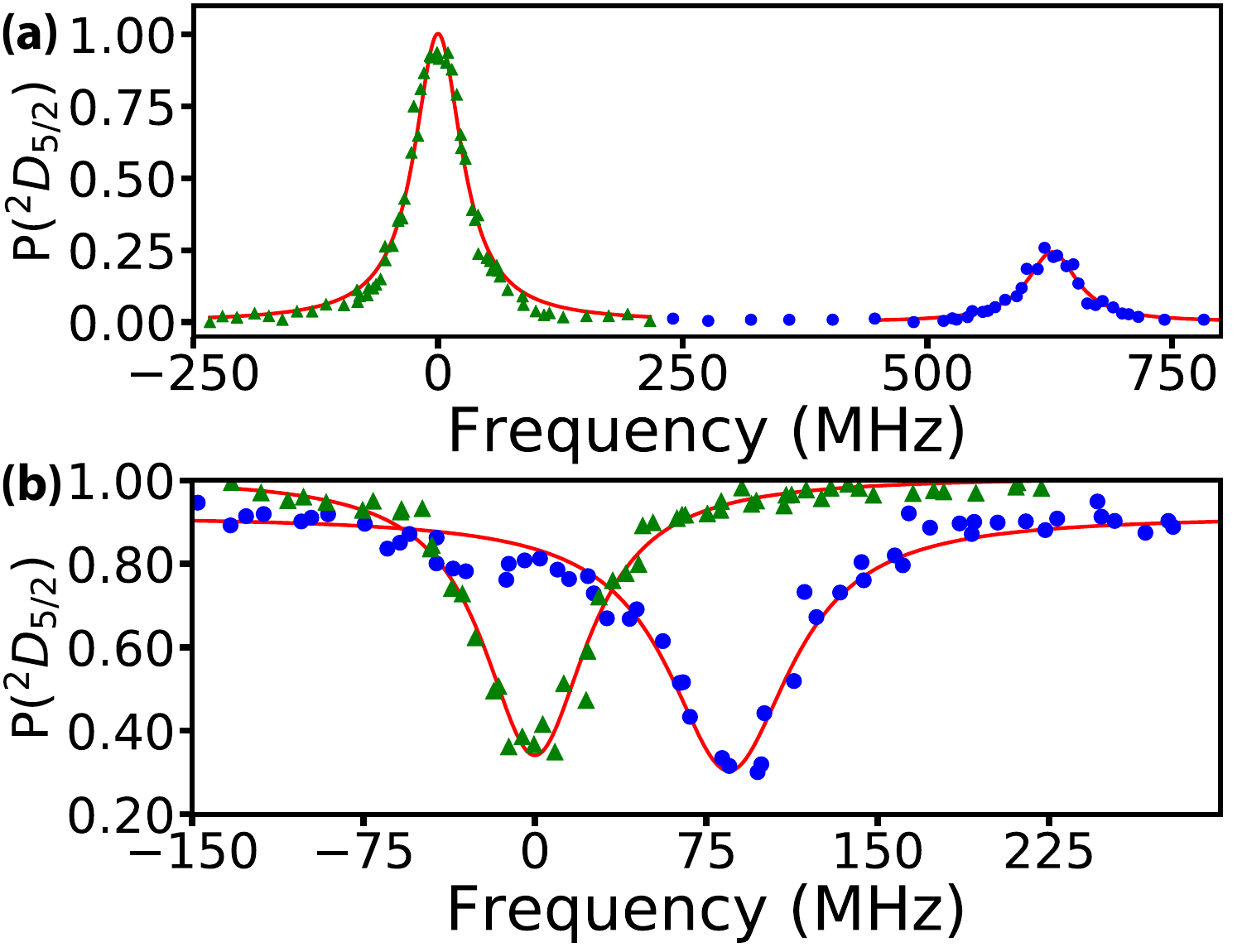}
\caption{\label{fig:spectroscopy}
(a) \DState\ population \emph{vs.} $\nu_{455}$ frequency. Triangles are data for the $\vert$\PState;$F\!\! =\!\! 2 \rangle  \leftrightarrow$ $\vert$\sState;$F\!\! =\!\! 1\rangle$ transition. Circles are data for the $\vert$\PState;$F\!\! =\!\! 1 \rangle  \leftrightarrow$ $\vert$\sState;$F\!\! =\!\! 1\rangle$ transition. 
Solid red line represents a fitted Lorentzian profile.
(b) \DState\ population \emph{vs.} $\nu_{614}$ frequency. Triangles are data for the $\vert$\PState;$F\!\! =\!\! 2 \rangle  \leftrightarrow$ $\vert$\DState;$F\!\! =\!\! 3\rangle$ transition. Circles are data for the $\vert$\PState;$F\!\! =\!\! 2 \rangle  \leftrightarrow$ $\vert$\DState;$F\!\! =\!\! 2\rangle$ transition.}
\end{figure}

Shelving of the \one\ qubit state via \DState $\leftrightarrow$ \sState \ transition requires application of a laser near 1762 nm ($\nu_{1762}$). 
Similarly, shelving of the \one\ state via optical pumping requires application of the frequencies $\nu_{455}$, $\nu_{585}$, and $\nu_{650}^c$ (and $\nu_{614}$ for deshelving). However, of these, only $\nu_{650}^c$ has been previously measured~\cite{hucul:2017}.
To determine these unknown frequencies we measure the  \PState $\leftrightarrow$ \sState \  and \PState $\leftrightarrow$ \DState \ isotope shifts relative to \natba\ ($\delta\nu_{138,133}^{455}$ and $\delta\nu_{138,133}^{614}$) and hyperfine splittings $\Delta_{3}$ and $\Delta_{5}$ (Fig. \ref{fig:ba_shelve}).
To measure $\Delta_3$ and $\delta\nu_{138,133}^{455}$, the atom is prepared in the $\vert$\sState;$F = 1\rangle$ manifold by optical pumping with $\nu_{650}^c$ and $\nu_{650}^{sb}$ after Doppler cooling. 
A tunable laser near 455~nm ($\nu_{455}$) is applied for 50 $\mu$s and its frequency scanned.
When the frequency is near one of the two allowed transitions, excitation followed by spontaneous emission from the \PState \ with branching ratios 0.74, 0.23, and 0.03 to the \sState, \DState, and \dState\ respectively~\cite{dutta:2016} optically pumps the ion to the \DState \ state.
The population remaining in the \sState\ and \dState\ states is then detected via collecting fluorescence while Doppler cooling and using threshold discrimination on the number of collected photons to decide if the atom was in the \DState\ state.
This sequence is repeated 200 times per laser frequency, and the average population is shown Fig. \ref{fig:spectroscopy}(a) as a function of frequency.
From these data, we find $\Delta_{3}$ = 623(30) MHz, and $\delta\nu_{138,133}^{455}$ = +358(30) MHz relative to \natba\!.

To measure $\Delta_5$ and $\delta\nu_{138,133}^{614}$, the atom is Doppler cooled, prepared in the $\vert$\sState;$F\!\!=\!\!1\rangle$ manifold, and shelved to the \DState \ state via one of the \PState \ hyperfine manifolds. 
The $\vert$\DState;$F\!\!=\!\!2\rangle$ manifold is prepared via shelving on the $\vert$\PState;$ F\!\! =\!\!1\rangle \leftarrow \vert$ \sState $; F\!\! =\!\! 1\rangle$ transition, as dipole selection rules forbid decay to the $\vert$\DState;$ F\!\! =\!\! 3\rangle$ state. 
Similarly, the $\vert$\DState;$F = 3\rangle$ manifold is prepared by shelving on the $\vert$\PState;$ F\!\! =\!\! 2\rangle\leftarrow \vert$\sState,$ F \!\!=\!\! 1\rangle$ transition, where 0.93 of decays to the \DState\ are to the $\vert$\DState;$F\!\! =\!\! 3\rangle$ manifold.
Next, a laser near 614~nm is applied and its frequency scanned.
When the frequency is near the $\vert$\PState$; F\!\! =\!\! 2\rangle  \leftrightarrow \vert$\DState$; F\!\! =\!\! 3 \rangle$ or $\vert$\PState$; F\!\! =\!\! 2\rangle  \leftrightarrow \vert$\DState$; F\!\! =\!\! 2 \rangle$ transition, spontaneous emission from the \PState\ state quickly deshelves the ion to the $\vert$\sState;$F\!\! =\!\! 1\rangle$ and \dState\ states. 
This deshelved population is then detected via Doppler cooling as shown in Figure \ref{fig:spectroscopy}(b). 
From these data, we find the \DState \ hyperfine splitting $\Delta_{5}$ = 83(30) MHz, and isotope shift   $\delta\nu_{138,133}^{614}$ = +216(30) MHz. 

With the required spectroscopy known, we can characterize the expected fidelity of optically-pumped electron shelving detection of the hyperfine qubit as follows. 
For SPAM of the $\ket{1}$ state, the initial state is prepared as described above and a laser resonant with the $\vert$\PState;$F\!\! =\!\! 2 \rangle \leftrightarrow$ $\vert$\sState;$F\!\! =\!\! 1\rangle$ transition ($\nu_{455}$) at an intensity below saturation is applied (Fig. \ref{fig:ba_shelve}).
After the first excitation of the atom, the \PState \ quickly ($\tau \approx$ 10 ns) spontaneously decays to either the \ \sState, \DState, or \dState \ with probabilities 0.74, 0.23, and 0.03 respectively.
Dipole selection rules forbid decay to the $\vert$\sState;$F\!\! =\!\! 0 \rangle$ (\zero\!) state, resulting in $\mathcal{F} = 0.88$ \ shelving fidelity, limited by population stranded in the \dState \ state. 
To further increase the shelving fidelity, a laser near 650 nm ($\nu_{650}^c$) resonant with the $\vert$\pState;$F\!\! =\!\! 0 \rangle \leftrightarrow$ $\vert$\dState;$F\!\! =\!\! 1\rangle$ states, and a laser near 585 nm ($\nu_{585}$)  resonant with the $\vert$\PState;$F\!\! =\!\! 2 \rangle \leftrightarrow$ $\vert$\dState;$F\!\! =\!\! 2\rangle$ states can be added.
Dipole selection rules forbid spontaneous emission to the $\vert$\sState;$F\!\! =\!\! 0 \rangle$ (\zero\!) state resulting in a fidelity of $\mathcal{F} \approx 0.999$.
This scheme is limited by off-resonant scatter of $\nu_{455}$ to the $\vert$\PState;$F\!\! =\!\! 1 \rangle$ state, which can spontaneously emit to the $\vert$\sState;$F\!\! =\!\! 0 \rangle$ state with a probability of 0.44.
If $\nu_{455}$ is linearly polarized parallel to the magnetic field direction ($\pi$-light), dipole selection rules forbid excitation from the $\vert$\PState;$F\!\! =\!\! 1;m_F\!\! =\!\! 0 \rangle  \leftrightarrow$ $\vert$\sState;$F\!\! =\!\!1;m_F\!\! = \!\!0\rangle$ for the first scattered photon, and the expected fidelity increases to $\mathcal{F} = 0.9998$.

For SPAM of the $\ket{0}$ state, initialization with optical pumping proceeds as described above. 
After preparation, the \one\ state is shelved as previously described, and the state is read out via Doppler cooling. 
During \one\ state shelving, off-resonant excitation to the $\vert$\PState$;F\!\! =\!\! 1\rangle$\ followed by spontaneous emission can shelve the ion to the \DState. 
This results in an expected SPAM fidelity of $\mathcal{F} = 0.9998$. 

\begin{figure}[t!]
\includegraphics[width=0.95\columnwidth]{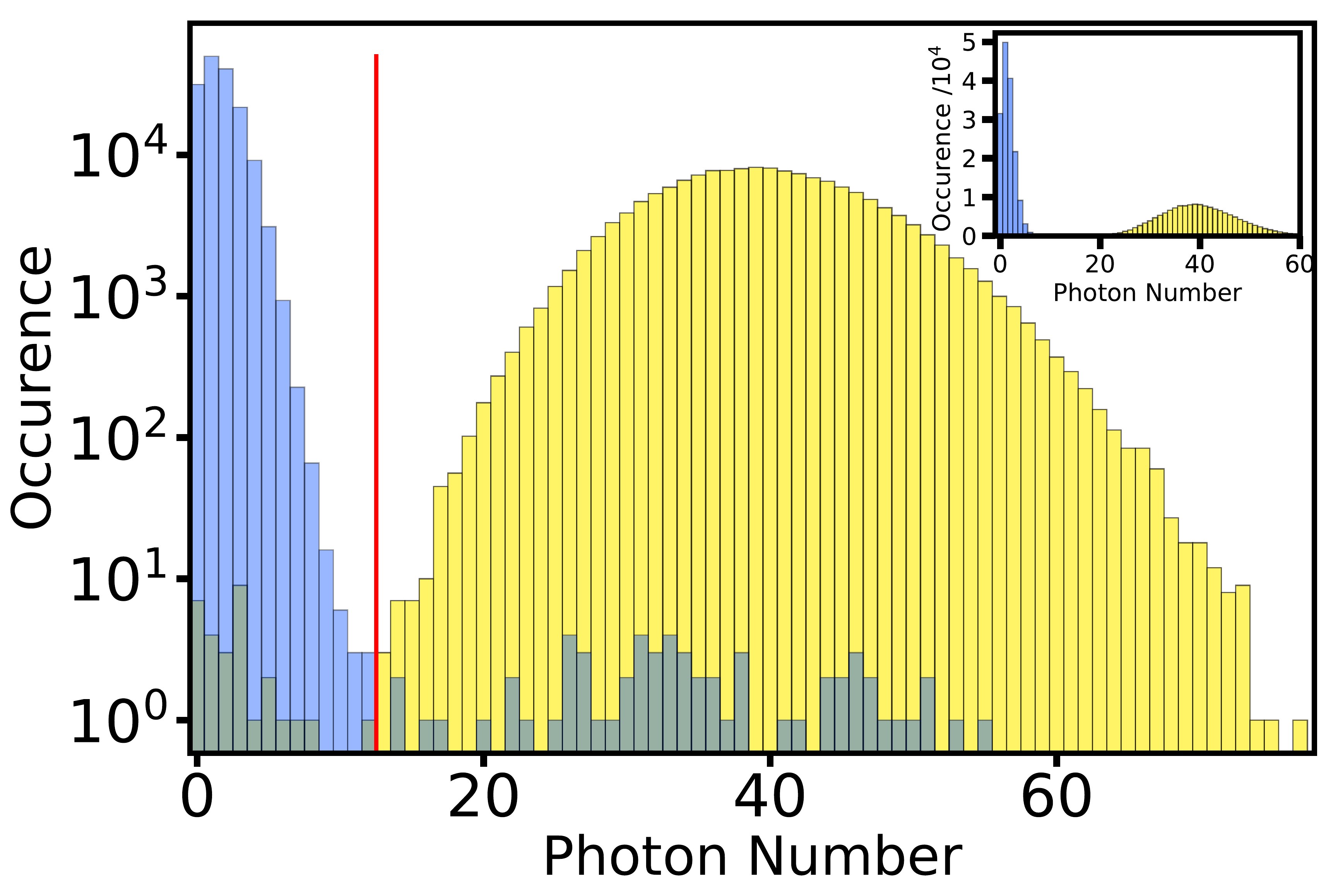}
\caption{\label{fig:log_histogram} We prepare and read out one of the two qubit states in blocks of 200 consecutive trials, alternating between qubit states for a total of 156,581 \zero\ state trials and 157,211 \one\ state trials. Detection of the \zero\ (bright) states returns an average of 39 collected photons, while detection of the \one\ (dark) state returns an average of 1 collected photon.
Using standard spin-1/2 techniques for \zero\ state preparation \cite{Acton2006, olmschenk:2007}, a five $\pi$-pulse composite pulse sequence \cite{ryan:2010} to prepare the \one\ state, and electron shelving for high fidelity readout, we measure an average SPAM error of $\epsilon$ = 2.9(6) x 10$^{-4}$.}
\end{figure}

To experimentally test these predictions, state preparation of each qubit state is applied to a single trapped \ba\ ion and read out using the optically-pumped shelving scheme. 
Each qubit state is attempted in blocks of 200 consecutive trials, followed by the other qubit state, for a combined total of 313,792 trials.
The number of photons detected after each experiment is plotted in Fig.~\ref{fig:log_histogram}, and a threshold at $n_{th} \leq 12$ photons is chosen based on the average number of counts from the bright state to discriminate between \zero and \one. 
The fraction of events in which an attempt to prepare the \zero \ state was measured to be \one\ is $\epsilon_{\vert 0\rangle} = 1.9(4) \times 10^{-4}$, while the fraction of experiments in which an attempt to prepare the \one \ state was measured to be \zero \ is $\epsilon_{\vert 1 \rangle} = 3.8(5) \times 10^{-4}$. 
The average SPAM fidelity is  $\mathcal{F} =1- \frac{1}{2}(\epsilon_{\vert 0\rangle} + \epsilon_{\vert 1 \rangle}) =  0.99971(6)$. 

Table~\ref{table:error} provides an error budget with estimates of the individual sources of error that comprise the observed infidelity. 
In addition to the previously discussed errors, we have experimentally determined several sources of infidelity.
The CP Robust 180 sequence is found to have an error of $\epsilon = 1 \times 10^{-4}$, determined by measuring the \one\ state SPAM infidelity as a function of the number of concatenated CP Robust 180 sequences.   
The state readout duration is determined by the need to statistically separate the \zero\ and \one\ state photon distributions.
Our limited numerical aperture requires detection for 4.5 ms, leading to an error due to spontaneous emission from the \DState\ state of 1 - exp($\frac{4.5\times 10^{-3}}{30}) \approx  1.5 \times 10^{-4}$. 
This could be corrected with maximum likelihood methods \cite{langer:2006, myerson:2008} or higher efficiency light collection \cite{noek:2013}. 
Finally, the readout of the \sState\ manifold is limited by background gas collisions, which we characterize by the preparation and readout fidelities of the \sState and \DState manifolds in \natba, for which we achieve $\mathcal{F} = 0.99997(2)$.

It should be possible to further improve this fidelity to $\mathcal{F}> 0.9999$ by use of a laser near 1762 nm (Fig. \ref{fig:ba_shelve}) in two ways.
First, optical-frequency qubit manipulations have been demonstrated (in other species) with a $\pi$-pulse fidelity of $\mathcal{F} = 0.99995$~\cite{gaebler:2016}.
Second, even without the narrow-band laser used for optical qubit manipulations, a 1762 nm laser could be used to saturate the transition and transfer 0.875 of the population into the \DState state. 
If this is followed with the optically-pumped shelving scheme, we expect an infidelity in state preparation of $\epsilon = 4 \times 10^{-5}$. 


\begin{table}[t!]\
\caption{Experimental error budget for state preparation and measurement (SPAM) of the \ba hyperfine qubit. Errors are estimates based on theoretical models and auxiliary experiments. The \zero\ state SPAM is limited by off-resonant scatter from the laser used for electron shelving. The \one\ state electron shelving is limited by the \PState\ hyperfine splitting, where off-resonant scatter can cause spontaneous emission to the \zero\ state. Spontaneous emission of the \DState\ state and preparation of the \one\ state via microwaves are the next largest contribution to the \one\ state SPAM error.}
\begin{ruledtabular}
\begin{tabular}{lc}
	Process                                         &Average error $\times 10^{-4}$ \\	
	\hline	 
	Initialization to \zero	                        &   0.1           		    \\
	\zero $\rightarrow$ \one CP Robust 180 sequence  & 0.5	                \\  
    Spontaneous decay during readout                & 0.7                    \\
    Shelving \one                                   & 1.0                    \\
    Off-resonant shelving \zero                     & 1.0                    \\
    Readout of \sState manifold                     & 0.1                    \\
    \hline
    Total average SPAM error                        & 3.4                    \\
\end{tabular}
\end{ruledtabular}
\label{table:error}
\end{table}


In summary, we report measurements in \ba of the  \PState \  and \DState \ hyperfine splittings  and  \PState $\leftrightarrow$ \sState \  and \PState $\leftrightarrow$ \DState \ transition frequencies, which are  required for high fidelity state readout and optical qubit manipulations. 
Using these measurements, we have demonstrated operation of the \ba hyperfine qubit, including use of the CP Robust 180 composite pulse sequence, to realize an average single-shot SPAM error of $\epsilon_s = 2.9(6) \times 10^{-4}$ via threshold discrimination. 
This represents a $\approx$ 2$\times$ reduction of SPAM error for any qubit \cite{harty:2014}, and is sufficient for single-shot, projective readout of a register of $\approx$ 2000 individually resolved qubits.

This work was supported by the US Army Research Office under award W911NF-18-1-0097. We thank Anthony Ransford, Christian Schneider and Conrad Roman for helpful discussions. We thank Peter Yu for technical assistance. 

\bibliography{BaHyperBib}

\end{document}